\hfuzz 2pt
\hbadness 10000
\font\titlefont=cmbx10 scaled\magstep1
\magnification=\magstep1

\null
\vskip 1.5cm
\centerline{\titlefont DISSIPATION AND DECOHERENCE}
\medskip
\centerline{\titlefont IN PHOTON INTERFEROMETRY}
\vskip 2.5cm
\centerline{\bf F. Benatti}
\smallskip
\centerline{Dipartimento di Fisica Teorica, Universit\`a di Trieste}
\centerline{Strada Costiera 11, 34014 Trieste, Italy}
\centerline{and}
\centerline{Istituto Nazionale di Fisica Nucleare, Sezione di
Trieste}
\vskip 1cm
\centerline{\bf R. Floreanini}
\smallskip
\centerline{Istituto Nazionale di Fisica Nucleare, Sezione di
Trieste}
\centerline{Dipartimento di Fisica Teorica, Universit\`a di Trieste}
\centerline{Strada Costiera 11, 34014 Trieste, Italy}
\vskip 2cm
\centerline{\bf Abstract}
\smallskip
\midinsert
\narrower\narrower\noindent
The propagation of polarized photons in optical media
can be effectively modeled by means of quantum dynamical semigroups.
These generalized time evolutions consistently describe phenomena leading
to loss of phase coherence and dissipation originating from the interaction
with a large, external environment. High sensitive experiments in the 
laboratory
can provide stringent bounds on the fundamental energy scale that characterizes
these non-standard effects.
\endinsert

\vfill\eject

{\bf 1. INTRODUCTION}
\medskip

Quantum systems are usually treated as isolated: their time evolution is
unitary, driven by the appropriate hamiltonian operator. In general however,
this should be regarded as an approximation: any external environment $\cal E$
unavoidably interacts with the system $\cal S$ under study,
making the resulting dynamics rather involved.[1-3]

The global system ${\cal S} + {\cal E}$ is closed, and its
time evolution is determined by the operator $e^{-i H_{\rm tot} t}$,
involving the total hamiltonian, that can be always decomposed as:
$$
H_{\rm tot}=H+H_{\cal E} + H'\ ,
\eqno(1.1)
$$
where $H$ drives the system $\cal S$ in absence of
$\cal E$, $H_{\cal E}$ describes the internal environment dynamics, while
$H'$ takes into account the interaction between $\cal S$ and $\cal E$.
Nevertheless, being interested only in the evolution of the subsystem $\cal S$
and not in the details of the dynamics of $\cal E$, one finally integrates
over the environment degrees of freedom. Describing
the states of subsystem and environment by appropriate density matrices,
the evolution in time of $\cal S$ will then be given by the transformation:
$$
\rho(0)\mapsto\rho(t)={\rm Tr}_{\cal E}\Big[ e^{-i H_{\rm tot} t}\
\rho(0)\otimes\rho_{\cal E}\ e^{i H_{\rm tot} t}\Big]\ ,
\eqno(1.2)
$$
where $\rho_{\cal E}$ describes the state of the environment at
$t=\,0$ (for simplicity, we assume $\cal S$ and $\cal E$ to be initially
uncorrelated; see [4] for a generalization).

The resulting map $\rho(0)\mapsto\rho(t)$ is rather complex,
involving in general non-linear and memory effects; it consistently describes
decoherence effects, leading to irreversibility and dissipation.
An explicit and mathematically precise description in terms of quantum
dynamical semigroups is however possible when the interaction between the
subsystem $\cal S$ and the environment is weak. These generalized 
time evolutions
are represented by linear transformations, mapping density matrices into
density matrices, while preserving very basic physical properties, like forward
in time composition law (semigroup property), entropy increase 
(irreversibility)
and complete positivity (that guarantees the correct physical interpretation
of the dynamics in all situations).[1-3]

Thanks to its generality and physical self-consistency, the description of
open systems in terms of quantum dynamical semigroups can be applied to model
the dynamics of any system in weak interaction with a large environment.[1-7]
In particular, it has recently been applied to describe possible effects
of irreversibility and dissipation induced by the evolution of strings
and branes. Indeed, quite in general the fundamental dynamics of these extended
objects gives rise at low energies to a weakly coupled environment, and
as a consequence to decoherence phenomena.[8, 9]

 From a more phenomenological point of view, similar effects have also been
described in the framework of quantum gravity: the quantum fluctuation of
the gravitational field and the appearance of virtual black holes make
space-time look ``foamy'' at distances comparable to Planck's length,
inducing non-standard phenomena leading
to possible loss of quantum coherence.[10-16]
Dissipation and decoherence are also the general result of the dynamics
in theories with large extra dimensions;[17] indeed, the possible 
energy leakage
into the bulk of space-time due to gravity effects
would inevitably inject noise into the boundary, thus inducing irreversibility
and dissipation at low energy in our brane-world.

Our present knowledge of string theory does not allow precise estimates of the
magnitude of these non-standard effects. Using a rough dimensional analysis,
one can nevertheless conclude that they should be rather small, being
suppressed by at least one inverse power of a large, fundamental mass 
scale (most
likely the Planck mass). Despite of this, they can be studied using 
interferometric
phenomena. Indeed, detailed investigations involving various 
elementary particle
systems (neutral mesons [18-22], neutrons [23], neutrinos [24, 25])
have shown that present and future experiments might soon reach
the required sensitivity in order to detect the new, non-standard phenomena.

This possibility looks particularly promising for photon interferometry
and more in general optical physics.[5-7] The sophistication of 
present laboratory
experiments in quantum optics is so high that decoherence effects induced by
a fundamental, ``stringy'' dynamics might be studied using available setups.

In the present work, we shall discuss in detail how these non-standard,
dissipative phenomena can affect the propagation of polarized photons immersed
in optically active media. A preliminary discussion has been presented in [26].
There, it has been shown that the dissipative phenomena manifest themselves
via depolarizing effects, that accumulate with time. Limits on the magnitude
of the parameters describing the new phenomena can therefore be obtained from
astrophysical and cosmological observations.[27]

In the following, different aspects of the quantum dynamical semigroup
description of photon propagation will be analyzed, focusing on the discussion
of possible laboratory tests.
As we shall see, the possibility of actually detecting the new,
dissipative effects are greatly enhanced by making them interfere
with those induced by time-dependent optical media.
For slowly varying media, the use of the adiabatic approximation is justified.
In this case, explicit expressions for relevant physical observables will
be given and discussed; the formulas can be used to fit actual experimental
data. The outcome of our investigation is that, at least in principle,
bounds on some of the parameters describing dissipation
and decoherence can be obtained using existing laboratory setups.

\vskip 1cm

{\bf 2. QUANTUM DYNAMICAL SEMIGROUPS}
\medskip

In describing the evolution of polarized photons we shall adopt the
standard effective description in terms of a two-dimensional Hilbert space,
the space of helicity states.[28-31]
A convenient basis in this space is given by the
circularly polarized states $|R\rangle$, $|L\rangle$. With respect to
this basis, any partially polarized photon state can be represented by a
$2\times2$ density matrix $\rho$, {\it i.e.} by an hermitian operator,
with positive eigenvalues and constant trace:
$$
\rho=\left[\matrix{\rho_1 & \rho_3\cr
                    \rho_4 & \rho_2}\right]\ ,
\qquad \rho_4=\rho_3^*\ .
\eqno(2.1)
$$

As explained in the Introduction, the evolution in time of $\rho$
will be described by mean of a quantum dynamical semigroup, {\it 
i.e.} by a linear
transformation generated by an equation of the following form:[1-3, 32-34]
$$
{\partial\rho(t)\over \partial t}= -i \Big[H(t)\, ,\ \rho(t)\Big]+ 
L[\rho(t)]\ .
\eqno(2.2)
$$
The first term in the r.h.s. is of hamiltonian form, while the piece
$L[\rho]$ takes into account the interaction with the external environment
and leads to irreversibility and dissipation.%
\footnote{$^\dagger$}{It should be noticed that in general the interaction with
the environment can also produce hamiltonian pieces in (2.2);[1-3, 9, 25]
however, in the present case these contributions can not be distinguished
from those originating from other birefringence phenomena.}

As mentioned in the introductory remarks, it is convenient to make the photons
cross an additional, non-dissipative, time-dependent optical medium,
whose properties can be suitably controlled. This will in general induce extra
birefringence effects on the polarized photons, and these can be conveniently
described in terms of a time-dependent, effective hamiltonian $H(t)$.
We shall assume a simple harmonic dependence on time:
$$
H(t)=\left[\matrix{
\omega_0+\mu & \nu\, e^{-i\lambda t}\cr
\nu\, e^{i\lambda t} & \omega_0-\mu}\right]\ ;
\eqno(2.3)
$$
this form is of sufficient generality for the considerations that follow.
In (2.3), the parameter $\omega_0$ represents the average photon 
energy, while the
real constants $\mu$ and $\nu$ induce the level-splitting
$\omega=(\mu^2+\nu^2)^{1/2}$ among the two instantaneous eigenstates.
As compared with the effects of this splitting, the dependence on time
of $H(t)$, characterized by the real frequency $\lambda$, will be assumed
to be slow: $\lambda\ll\omega$; this is the situation
that is most likely to be reproduced by actual laboratory setups.

The additional piece $L[\rho]$ in the evolution equation (2.2) is not
of hamiltonian form, and induces a mixing-enhancing mechanism leading in
general to irreversibility and loss of quantum coherence.
In order to write it down explicitly, it is useful to adopt a vector-like
notation and collect the entries $\rho_1$, $\rho_2$, $\rho_3$, $\rho_4$
of the density matrix (2.1) as the components of the four-dimensional abstract
vector $|\rho\rangle$. The evolution equation (2.2) can then be rewritten
as a Schr\"odinger (or diffusion) equation:
$$
{\partial\over\partial t} |\rho(t)\rangle=
\Big[ {\cal H}(t) + {\cal L}\Big]\, |\rho(t)\rangle\ ,
\eqno(2.4)
$$
where the $4\times4$ matrix $\cal H$ takes into account the hamiltonian
contributions,
$$
{\cal H}(t)=i\,\left[
\matrix{0 & 0 & \nu\, e^{i\lambda t} & -\nu\, e^{-i\lambda t}\cr
         0 & 0 & -\nu\, e^{i\lambda t} & \nu\, e^{-i\lambda t}\cr
		\nu\, e^{-i\lambda t} & -\nu\, e^{-i\lambda t} & -2\mu & 0\cr
		-\nu\, e^{i\lambda t} & \nu\, e^{i\lambda t} & 0 & 
2\mu}\right]\ ,
\eqno(2.5)
$$
while the dissipative part $\cal L$ can be fully parametrized in terms of six
real constants $a$, $b$, $c$, $\alpha$, $\beta$, and $\gamma$, as 
follows:[1, 18]
$$
{\cal L}=\left[\matrix{ -D\phantom{^*} & \phantom{-}D\phantom{^*} &
-C\phantom{^*} & -C^*\cr
                          \phantom{-}D\phantom{^*} & -D\phantom{^*} &
\phantom{-}C\phantom{^*} & \phantom{-}C^*\cr
						-C^* & \phantom{-}C^* 
& -A\phantom{^*} &
\phantom{-}B\phantom{^*}\cr
						-C\phantom{^*} & 
\phantom{-}C\phantom{^*} &
\phantom{-}B^* & -A\phantom{^*}}\right]\ ,
\eqno(2.6)
$$
where for later convenience the combinations:
$$
A=\alpha+a\ ,\quad B=\alpha -a +2ib\ ,\quad C=c+i\beta\ ,\quad D=\gamma\ ,
\eqno(2.7)
$$
have been introduced. The six parameters are not all independent:
they need to satisfy the following inequalities:[1-3, 18, 35]
$$
\eqalign{
&2\,R\equiv\alpha+\gamma-a\geq0\ ,\cr
&2\,S\equiv a+\gamma-\alpha\geq0\ ,\cr
&2\,T\equiv a+\alpha-\gamma\geq0\ ,\cr
&RST-2\, bc\beta-R\beta^2-S c^2-T b^2\geq 0\ .
}\hskip -1cm
\eqalign{
&RS-b^2\geq 0\ ,\cr
&RT-c^2\geq 0\ ,\cr
&ST-\beta^2\geq 0\ ,\cr
&\phantom{\beta^2}\cr
}\qquad\qquad
\eqalign{
&a\geq0\ ,\cr
&\alpha\geq0\ ,\cr
&\gamma\geq0\ ,\cr
&\phantom{\beta^2}\cr
}
\eqno(2.8)
$$
These relations are the consequence of the property of complete positivity
that assures the correct physical interpretation of the time evolution
$|\rho(0)\rangle\to |\rho(t)\rangle$ generated by (2.4) in all situations;
without this condition, serious inconsistencies in general arise
(for more details, see [35]).

The effective environment generated by the fundamental ``stringy'' dynamics
can be considered to be in thermal equilibrium;[9] the decoherence effects
induced on the photons are therefore stationary, so that the six
parameters $a$, $b$, $c$, $\alpha$, $\beta$, $\gamma$ in (2.6), (2.7)
can be taken to be time-independent.
Nevertheless, let us mention that the evolution equation (2.2) can be 
generalized
to take into account non-stationary dissipative contributions:
these would typically arise for environments that are out of equilibrium,
giving rise in general to time-dependent intractions with the photons.[32-34]
On the other hand, a physically consistent, general
formulation of non-linear dissipative dynamics is not yet available.

Once the evolution equation (2.4) is solved, one can easily compute 
any physical
property involving polarized photons. Indeed, in the formalism of 
density matrices,
any observable $\cal O$ is represented by an hermitian matrix, that can be
decomposed as in (2.1). The evolution in time of its mean value is 
then obtained
by taking its trace with the density operator $\rho(t)$:
$$
\langle{\cal O}(t)\rangle\equiv{\rm Tr}\Big[{\cal O}\,\rho(t)\Big]=
{\cal O}_1\, \rho_1(t) + {\cal O}_2\, \rho_2(t) + {\cal O}_3\, \rho_4(t)
+ {\cal O}_4\, \rho_3(t)\equiv \langle{\cal O} | \rho(t)\rangle\ .
\eqno(2.9)
$$

In the case of photons, of particular interest is the observable that 
correspond
to a fully polarized state, identified by the two angles $\theta$ and 
$\varphi$;
it is explicitly given by the following projector operator
$$
{\cal O}_{\theta,\varphi}={1\over2}\left[\matrix{
1+\sin\varphi\sin2\theta & \cos2\theta-i\cos\varphi\sin2\theta\cr
\cos2\theta+i\cos\varphi\sin2\theta & 1-\sin\varphi\sin2\theta\cr}\right]\ .
\eqno(2.10)
$$
Its mean value gives the probability ${\cal P}_{\theta,\varphi}(t)$ that the
evolved state $|\rho(t)\rangle$ be found at time $t$ in the polarization state
determined by $\theta$ and $\varphi$; it is proportional to the intensity
curve that can be detected at an appropriate interferometric apparatus.

\vfill\eject

{\bf 3. TRANSITION PROBABILITIES}
\medskip

In order to find explicit solutions of the evolution equation (2.4),
it is convenient to perform a time-dependent unitary transformation
and study it in the basis of instantaneous eigenvectors
$|v^{(\pm)}(t)\rangle$ of the hamiltonian (2.3),
$H(t)\, |v^{(\pm)}(t)\rangle=(\omega_0\pm\omega)\, |v^{(\pm)}(t)\rangle$;
using the four-vector notation, one then writes
$$
|\widetilde\rho(t)\rangle={\cal U}(t)\, |\rho(t)\rangle\ ,
\eqno(3.1)
$$
where
$$
{\cal U}(t)={1\over 2\omega}\,\left[\matrix{
\omega+\mu & \omega-\mu & \nu\, e^{i\lambda t} & \nu\, e^{-i\lambda t}\cr
\omega-\mu & \omega+\mu & -\nu\, e^{i\lambda t} & -\nu\, e^{-i\lambda t}\cr
-\nu\, e^{-i\lambda t} & \nu\, e^{-i\lambda t} & \omega +\mu &
-(\omega-\mu) e^{-2i\lambda t}\cr
-\nu\, e^{i\lambda t} & \nu\, e^{i\lambda t} & -(\omega-\mu) e^{2i\lambda t} &
\omega+\mu }\right]\ .
\eqno(3.2)
$$
In the new basis, the hamiltonian contribution in (2.4) becomes diagonal:
$$
\widetilde{\cal H}={\cal U}(t)\,{\cal H}(t)\, {\cal U}^\dagger (t)
={\rm diag}[0,0,-2i\omega, 2i\omega]\ ;
\eqno(3.3)
$$
the four entries coincide with the eigenvalues of the operator 
$-i[H(t), \cdot\,]$,
and therefore are given by the differences of the eigenvalues
$\omega_0\pm\omega$ of $H(t)$. However, since ${\cal U}(t)$ is time dependent,
the evolution equation for the transformed vector $|\widetilde\rho(t)\rangle$
involves an effective hamiltonian:
$$
{\partial\over\partial t} |\widetilde\rho(t)\rangle=
\Big[ {\cal H}_{\rm eff}(t) + \widetilde{\cal L}(t)\Big]\,
|\widetilde\rho(t)\rangle
\ ,
\eqno(3.4)
$$
with
$$
{\cal H}_{\rm eff}(t)=\widetilde{\cal H}
+ \dot{\cal U}(t)\, {\cal U}^\dagger(t)\ ;
\eqno(3.5)
$$
further, the dissipative contribution becomes time-dependent:
$$
\widetilde{\cal L}(t)={\cal U}(t)\,{\cal L}\ {\cal U}^\dagger (t)\ .
\eqno(3.6)
$$
One can check that its explicit form is as in (2.6), with the new parameters
$\widetilde A$, $\widetilde B$, $\widetilde C$, $\widetilde D$ linear 
combinations
of the old ones $A$, $B$, $C$, $D$:%
\footnote{$^\dagger$}{This is a general property of any quantum dynamical
semigroup, whose explicit form is in fact basis-independent.[1-3]}
$$
\eqalignno{
&\widetilde A=A +{\nu^2\over 2\omega^2}\bigg[2D-A+
{\cal R}e\Big(B e^{2i\lambda t}\Big)\bigg]-{2\mu\nu\over\omega^2}
{\cal R}e\Big(C e^{-i\lambda t}\Big)\ ,&(3.7a)\cr
&\widetilde B=e^{-2i\lambda t}\bigg\{\bigg(1-{\nu^2\over2\omega^2}\bigg)
{\cal R}e\Big(B e^{2i\lambda t}\Big)
+ {i\mu\over\omega}{\cal I}m\Big(B e^{2i\lambda t}\Big)
+ {2\mu\nu\over\omega^2}{\cal R}e\Big(C e^{-i\lambda t}\Big)\cr
&\hskip 6cm - {2i\nu\over\omega}{\cal I}m\Big(C e^{-i\lambda t}\Big)
-{\nu^2\over2\omega^2}\Big(2D-A\Big)\bigg\}\ ,&(3.7b)\cr
&\widetilde C=e^{i\lambda t}\bigg\{\bigg(1-{2\nu^2\over\omega^2}\bigg)
{\cal R}e\Big(C e^{-i\lambda t}\Big)
+ {i\mu\over\omega}{\cal I}m\Big(C e^{-i\lambda t}\Big)
- {\mu\nu\over2\omega^2}\bigg[2D-A+{\cal R}e\Big(B e^{2i\lambda 
t}\Big)\bigg]\cr
&\hskip 6cm + {i\nu\over2\omega}{\cal I}m\Big(B e^{2i\lambda t}\Big)\bigg\}\ ,
&(3.7c)\cr
&\widetilde D= D-{\nu^2\over2\omega^2}\bigg[2D-A+
{\cal R}e\Big(B e^{2i\lambda t}\Big)\bigg]+{2\mu\nu\over\omega^2}
{\cal R}e\Big(C e^{-i\lambda t}\Big)\ .&(3.7d)\cr
}
$$

When the system hamiltonian $H(t)$ is slowly varying, the explicit
dependence on time of ${\cal H}_{\rm eff}(t)$ is very mild, so that the
adiabatic approximation can be used in studying (3.4). In general, this
is justified when the transitions induced by the explicit time dependence of
the hamiltonian are suppressed with respect to its natural level splitting.[36]
In the present case, this condition is guaranteed by the starting assumption:
$\lambda\ll\omega$. Within this approximation,
one can neglect the off-diagonal terms in
the contribution $\dot{\cal U}(t)\, {\cal U}^\dagger(t)$, so that
${\cal H}_{\rm eff}$ becomes diagonal:
$$
{\cal H}_{\rm eff}={\rm diag}\Big[0,0,-2i(\omega+\lambda_B), 
2i(\omega+\lambda_B)
\Big]\ .
\eqno(3.8)
$$
(As explained in the Appendix, in the case of the hamiltonian (2.3) this
result can be directly checked.%
\footnote{$^\dagger$}{Indeed, the evolution
generated by the hamiltonian $H(t)$ in (2.3) can be written in closed form.
In general however, this is no longer possible when the
dissipative contribution in (2.2) is non-vanishing.}
\hskip -.3cm)
The additional phase contribution $\lambda_B$ to the
finite-time evolution operator $e^{{\cal H}_{\rm eff} t}$ has a 
precise physical
meaning: it gives the Berry phase that in general accumulates with 
time;[37, 38]
indeed, one easily checks that:
$$
\lambda_B={\lambda\over2}\bigg(1-{\mu\over\omega}\bigg)
\equiv \mp\, i\, \langle v^{(\pm)}(t)|\,{\partial\over\partial t}\,
|v^{(\pm)}(t)\rangle\ .
\eqno(3.9)
$$
Being encoded in the diagonal part of $\dot{\cal U}(t)\, {\cal U}^\dagger(t)$,
Berry's contribution is directly connected to the characteristic 
properties of the
starting hamiltonian $H(t)$, and not to the use of the adiabatic approximation.

In absence of the dissipative piece, $\widetilde{\cal L}=\,0$, the evolution
in time of any given initial state $|\rho(0)\rangle$ can then be written as:
$$
|\rho(t)\rangle={\cal U}^\dagger(t)\cdot {\cal M}_0(t)\cdot {\cal U}(0)\,
|\rho(0)\rangle\ ,\qquad {\cal M}_0(t)=e^{{\cal H}_{\rm eff} t}\ .
\eqno(3.10)
$$
Using this expression, one can compute the evolution of physically
relevant observables, and in particular transition probabilities.
An experimentally relevant example is given by the probability
${\cal P}_\theta(t)$ of finding an initially left-polarized photon in a state
with linear polarization along the direction $\theta$ at time $t$.
Using the general definition (2.9) and the expression in (2.10) with 
$\varphi=\,0$,
from the evolution map (3.10) one explicitly finds:
$$
\eqalign{
{\cal 
P}_\theta(t)={1\over2}\bigg\{1+{\mu\nu\over\omega^2}\cos(2\theta-\lambda 
t)
\Big[\cos &\big[2(\omega+\lambda_B)t-\lambda t\big]-1\Big]\cr
&+{\nu\over\omega}\sin(2\theta-\lambda t)
\sin[2(\omega+\lambda_B)t-\lambda t\big]\bigg\}\ .}
\eqno(3.11)
$$
This expression further simplifies for a vanishingly small $\mu$;
in this case, it can be conveniently rewritten as:
$$
{\cal P}_\theta(t)={1\over2}\bigg\{1+{1\over2}\Big[\cos(2\omega 
t+\lambda t-2\theta)
+\cos(2\omega t-\lambda t+2\theta+\pi)\Big]\bigg\}\ .
\eqno(3.12)
$$

This intensity pattern can be studied by means of an interferometric setup:
one can then extract amplitudes and phases
of the various Fourier components that characterize the probability (3.12).
In particular, the presence of the modulation
$e^{-i\lambda t}$ in the hamiltonian (2.3) describing
in the optical medium crossed by the photon beam
leads to a symmetric shift of the fundamental birefringence frequency $2\omega$
by the small amount $\lambda$.
Notice that this result is a consequence of the presence of Berry's phase
contribution, that now takes the simplified expression $\lambda_B=\lambda/2$;
indeed, neglecting this contribution
would have produced an asymmetric split of the fundamental frequency.
An experimental analysis of the intensity pattern (3.12) can then 
allow a direct
identification of Berry's phase.

To see how this description is modified by the presence of 
dissipative phenomena,
one needs to study the evolution equation (3.4)
with a non-vanishing $\widetilde{\cal L}$.
Although in general the effects induced by the interaction with the environment
are parametrized by the six real constants $a$, $b$, $c$, $\alpha$, $\beta$,
and $\gamma$, there are physically motivated
instances for which only one of them is actually
non-zero. For example, this happens when $\gamma$ is vanishingly small;
in this case, the inequalities (2.8) further imply $a=\alpha$ and
$b=c=\beta=\,0$.%
\footnote{$^\dagger$}{There are essentially two known ways
of implementing the condition of weak interaction between subsystem
and environment:[1-3] the singular coupling
limit (in which the time-correlations in the environment are assumed to be much
smaller than the typical time scale of the subsystem)
and the weak coupling limit (in which it is the subsystem
characteristic time scale that becomes large).
One can check that the second situation
leads precisely to the condition $\gamma=\,0$.[9, 25]}
In this case, the entries of the matrix $\widetilde{\cal L}$ are all
proportional to $\alpha$, and assuming as before $\mu=\,0$,
from (3.7) one explicitly obtains:
$$
\widetilde A=\widetilde D=\alpha\ ,\qquad \widetilde B=\alpha\, 
e^{-2i\lambda t}\ ,
\qquad \widetilde C=\,0\ .
\eqno(3.13)
$$
Although the resulting expression for $\widetilde{\cal L}$ is still explicitly
time-dependent, the evolution equation (3.4) can be exactly integrated;
one finds:
$$
|\widetilde\rho(t)\rangle={\cal M}(t)\, |\widetilde\rho(0)\rangle\ 
,\qquad\qquad
{\cal M}(t)=e^{-\alpha t}\ \left[\matrix{\Theta(t) & 0\cr
                                               0 & \Xi(t)}\right]\ ,
\eqno(3.14)
$$
where the $2\times2$ matrices $\Theta(t)$ and $\Xi(t)$ can be expressed
in terms of the Pauli matrices $\sigma_1$, $\sigma_3$
and the identity $\sigma_0$:
$$
\Theta(t)=e^{\alpha t \sigma_1}\ ,
\qquad
\Xi(t)=e^{-i\lambda t\, \sigma_3}\ \bigg[
\cos2\Omega t\ \sigma_0 -{i\omega\over\Omega}\sin2\Omega t\ \sigma_3
+{\alpha\over2\Omega}\sin2\Omega t\ \sigma_1\bigg]\ ,
\eqno(3.15)
$$
and
$$
\Omega=\sqrt{\omega^2-\alpha^2/4}\ .
\eqno(3.16)
$$
Using the expression of the evolution matrix ${\cal M}(t)$ above in place
of ${\cal M}_0(t)$ in (3.10), one finally obtains the dynamical map
$|\rho(0)\rangle\to|\rho(t)\rangle$ in presence of dissipative effects.
Accordingly, the expressions of physically interesting observables change.
In particular,
the transition probability ${\cal P}_\theta(t)$ of finding an initial
circularly polarized photon in a linearly polarized state at time $t$
becomes:
$$
{\cal P}_\theta(t)={1\over2}\bigg\{1+{\omega\over2\Omega}e^{-\alpha t}
\Big[\cos\big(2\Omega t+\lambda t-2\theta\big)
+\cos\big(2\Omega t-\lambda t+2\theta+\pi\big)\Big]\bigg\}\ .
\eqno(3.17)
$$
The presence of dissipation affects the expression of ${\cal P}_\theta(t)$
through the introduction of the exponential damping term together with
the amplitude rescaling by the factor $\omega/\Omega$, and the change in the
birefringence frequency from $\omega$ to $\Omega$.
On the other hand, note that the symmetric shift in frequency by
the amount $\lambda$ induced by Berry's phase contribution remains unchanged.
This is not surprising: the geometrical
mechanism leading to the presence of Berry's phase is completely different
from the physical phenomena leading to irreversibility and dissipation,
and this fact is clearly reflected in the expression of the 
transition probability
(3.17). As a result, the dissipative contributions
and those originating from Berry's phase can be independently probed.

Similarly to the expression in (3.12), also the intensity pattern described
by (3.17) can be, at least in principle, experimentally studied
using Fourier analysis. Notice however that the oscillatory behaviour
in (3.17) critically depends on the magnitude of the non-standard effects
induced by the presence of the environment;
indeed, for sufficiently large $\alpha$, the frequency $\Omega$
becomes purely imaginary, so that the only remaining harmonic dependence
in ${\cal P}_\theta(t)$ is driven by the small frequency $\lambda$:
$$
{\cal P}_\theta(t)={1\over2}\bigg\{1+{\omega\over\Omega}e^{-\alpha t}
\sinh(\Omega t)\, \sin(2\theta-\lambda t)\bigg\}\ .
\eqno(3.18)
$$

In any case, independently from the relative magnitude of $\alpha$
and $\omega$, the damping effects always prevail for large times:
in this limit, the transition probability
${\cal P}_\theta(t)$ takes the constant value $1/2$.
One can show that this result is independent from the approximation
used to derive (3.17). Actually, in presence of dissipative phenomena
all transition probabilities asymptotically tend to constant values,
corresponding to the transition to a totally depolarized state.[39, 26]

A different treatment is possible when
the non-standard parameters $a$, $b$, $c$, $\alpha$, $\beta$, and $\gamma$
can be considered to be small in comparison with the characteristic
system energy $\omega$. This is likely to be the case in most 
standard laboratory
situations: indeed, the main source of birefringence
effects is usually the propagation in laboratory controlled
optical media, and not the weak
interaction with an external environment. In this case, the additional
piece $\widetilde{\cal L}$ in (3.4) can be treated as a perturbation, and the
evolution matrix ${\cal M}(t)$ in (3.14) can thus be expressed as the following
series expansion:
$$
\eqalign{{\cal M}(t) =\, & e^{{\cal H}_{\rm eff} t} \bigg\{ 1
+\int_0^t dt_1\, e^{-{\cal H}_{\rm eff} t_1}\, \widetilde{\cal L}(t_1)\,
e^{{\cal H}_{\rm eff} t_1}\cr
&\hskip 1cm +\int_0^t dt_1 \int_0^{t_1} dt_2\, e^{-{\cal H}_{\rm eff}t_1}\,
\widetilde{\cal L}(t_1)\, e^{{\cal H}_{\rm eff} (t_1-t_2)}\,
\widetilde{\cal L}(t_2)\, e^{{\cal H}_{\rm eff} t_2}
+\dots\bigg\}\ .}
\eqno(3.19)
$$

Useful information on the presence of dissipative effects can already 
be obtained
by considering only first order terms in the small parameters. Within this
approximation, the transition probability ${\cal P}_\theta(t)$ takes 
the following
explicit form:
$$
\eqalign{
{\cal P}_\theta(t)={1\over2} &+ {e^{-(D+A/2)t}\over2}\bigg\{
-\Delta(t)\, \cos(2\theta-\lambda t)\cr
&+\bigg[\bigg( 1+{|B|\over2\lambda}\sin\lambda t\,\sin(\lambda t+\phi_B)\bigg)
\sin2\omega t - \Phi(t)\bigg]\sin(2\theta-\lambda t)\bigg\}\ ,}
\eqno(3.20)
$$
where
$$
\eqalignno{
&\Delta(t)={|C|\over2}\bigg[ {2\lambda\over 4\omega^2-\lambda^2}\sin\phi_C
-{\sin(2\omega t+\lambda t-\phi_C)\over2\omega+\lambda}
-{\sin(2\omega t-\lambda t+\phi_C)\over2\omega-\lambda}\bigg]\cr
&\hskip 1cm +{|B|\over8}\bigg[{2\omega\over\omega^2-\lambda^2}\sin\phi_B
+{\sin(2\omega t-2\lambda t-\phi_B)\over \omega-\lambda}
-{\sin(2\omega t+2\lambda t+\phi_B)\over\omega+\lambda}\bigg]\ ,\hskip 1cm
&(3.21a)\cr
&\null\cr
&\Phi(t)={|B|\over4}\sin(\lambda t+\phi_B)\bigg[
{\sin(2\omega+\lambda)t\over2\omega+\lambda}
-{\sin(2\omega -\lambda)t\over2\omega-\lambda}\bigg]\cr
&\hskip 3cm +2|C|\sin(\lambda t/2-\phi_C)\bigg[
{\sin(2\omega-\lambda/2)t\over4\omega-\lambda}
+{\sin(2\omega+\lambda/2)t\over4\omega+\lambda}\bigg]\ ,&(3.21b)\cr
}
$$
while $\phi_B$ and $\phi_C$ are the phases of $B$ and $C$, the combination
of dissipative parameters introduced in (2.7). In writing (3.20) we have
reconstructed the exponential damping factor by putting together
terms linear in $t$: this is consistent at the used level of approximation;
the large time asymptotic behaviour of ${\cal P}_\theta(t)$ mentioned before
is thus reproduced. The expression in (3.20)
is clearly much more involved than the ones presented before:
it represents the most
general form that the transition probability ${\cal P}_\theta(t)$
can take in presence of small dissipative effects.

\vskip 1cm

{\bf 4. DISCUSSION}
\medskip

The propagation of polarized photons in optical media can be
consistently discussed within the formalism of open quantum systems,
{\it i.e.} as a system in interaction with a large environment.
This treatment can be physically justified in the framework
of string and brane theory, whose dynamics can be effectively described
at low energies as a weakly coupled environment, inducing
non-standard phenomena leading in general
to decoherence and dissipation. Quantum dynamical semigroups give a
physically consistent and mathematically precise description of these
non-standard effects; it turns out that they can be fully parametrized
in terms of the phenomenological constants $a$, $b$, $c$, $\alpha$,
$\beta$, and $\gamma$ introduced in (2.6), (2.7).

As discussed in the previous section,
the presence of these constants modify in a distinctive way the time
evolution of physically interesting observables, that can be experimentally
studied using interferometric setups.
In particular, the new, dissipative phenomena manifest themselves
through depolarizing effects, via the presence of exponential damping factors,
and suitable shifts in the frequencies describing birefringence effects.

Although a detailed discussion on possible devices that can be used to measure
such effects is surely beyond the scope of the present investigation,
some general considerations can nevertheless be given.%
\footnote{$^\dagger$}{An additional discussion, although referred to the
analysis of a specific experimental apparatus, can be found in Ref.[40].}
Recalling for instance the expression (3.17) for the transition
probability ${\cal P}_\theta(t)$, one immediately realizes that the 
possibility of
detecting the depolarizing effects induced by the non-standard,
dissipative phenomena is connected with the ability of isolating
and extracting the exponential
factor $e^{-\alpha t}$ from the experimental data. The sensitivity of this
measure clearly increases with $t$, so that large optical paths are in general
required. This can be achieved by using high quality optical cavities.
By adjusting in a controlled way the ``finesse'' and the optical properties
of the cavity, one should be able to reconstruct from the measured 
signal the time
(or path-length) dependence of the probability ${\cal P}_\theta(t)$,
and therefore extract information on the dissipative parameters
both from the damping factors and the oscillating terms.

The actual visibility of these parameters clearly depends on
their magnitude. A precise a priori evaluation
would require a detailed knowledge of string
theory; nevertheless, an order of magnitude estimate can be obtained
using the general theory of open systems. Indeed, quite in general the
dissipative effects induced by the weak interaction with an external
environment can be roughly evaluated to be at most proportional to the
square of the typical energy scale of the system, while suppressed by an
inverse power of the characteristic energy scale of the environment.

In the case of polarized photons, the system energy coincides with the average
photon energy $\omega_0$, while the typical energy scale of the environment
coincides with the mass $M_F$ that characterizes the fundamental, underlying
dynamics. As a consequence, the values of the parameters
$a$, $b$, $c$, $\alpha$, $\beta$, and $\gamma$ can be predicted to be roughly
of order $\omega_0^2/M_F$.

In the case of laboratory experiments using ordinary laser beams, the photon
energy $\omega_0$ is fixed; therefore the expected
magnitude of the new, non-standard effects is determined by the value of $M_F$.
This fundamental scale can be as large as the Planck mass, but can also be
considerably smaller in models of large extra dimensions. Fortunately,
as stressed before, the description of decoherence phenomena by means
of quantum dynamical semigroups is very general, and quite independent from the
actual microscopic mechanism responsible for the appearance of the new effects.
As a result, an experimental study of the transition probabilities discussed
in the previous section can give model-independent
indications of the presence of the non-standard, dissipative phenomena.
In turn, this would allow the derivation of
interesting bounds on the magnitude of the
fundamental scale $M_F$, thus providing useful information on the
underlying ``stringy'' dynamics.

The analysis of the previous sections have been limited to the study
of dissipative evolutions for polarization states of a single photon.
The whole treatment can be naively extended by linearity to include also
the case of multi-photon states. Generalizing the one-photon dissipative
dynamics generated by (2.2) to the case of multi-photon states is however
not completely straightforward. Indeed, the photons obey the Bose statistics,
and this property should be preserved by the time-evolution.
It turns out that physically\break
\eject
\noindent
acceptable multi-photon dissipative dynamics
can not be simply expressed as the product of single-photon time-evolutions:
a more refined treatment is necessary (see [41] for details).
This fact might have interesting consequences in various aspects
of quantum optics.

\vskip 2cm

{\bf APPENDIX}
\medskip

As mentioned in the text, the evolution flow generated by the hamiltonian
$H(t)$ in (2.3) can be exactly integrated. The explicit expression for the
corresponding unitary evolution operator $U(t)$ is given by:
$$
U(t)=e^{-i\lambda t \sigma_3/2}\ \bigg[
\cos\Omega_0 t\ \sigma_0
-i\bigg({2\mu-\lambda\over2\Omega_0}\bigg)\sin\Omega_0 t\ \sigma_3
-{i\nu\over\Omega_0}\sin\Omega_0 t\ \sigma_1\bigg]\ ,
\eqno(A.1)
$$
where $\Omega_0=\big[(\mu-\lambda/2)^2+\nu^2\big]^{1/2}$, while $\sigma_1$
and $\sigma_3$ are Pauli matrices and $\sigma_0$ the identity.
Indeed, one easily verifies that:
$$
\dot U(t)=-i H(t)\, U(t)\ ,\qquad U(0)=\sigma_0\ .
\eqno(A.2)
$$

Having the explicit solution of $(A.2)$, one can now check directly the
correctness of the adiabatic approximation used in Section 3.
To this purpose, one needs to consider the appropriate evolution operator
$\widetilde U(t)$ in the basis of the instantaneous eigenvalues of the
hamiltonian $H(t)$. The change of basis is provided by the transformation
matrix
$$
T(t)={1\over\sqrt{2\omega(\mu+\omega)}}\left[\matrix{
\mu+\omega & -\nu\, e^{-i\lambda t}\cr
\nu\, e^{i\lambda t} & \mu+\omega}\right]\ .
\eqno(A.3)
$$
In the limit of small $\lambda$, one then easily verifies that the 
new evolution
operator:
$$
\widetilde U(t)= T^\dagger(t)\, U(t)\, T(0)\ ,
\eqno(A.4)
$$
indeed becomes diagonal:
$$
\widetilde U(t)=\left[\matrix{e^{-i(\omega+\lambda_B) t} & 0\cr
                                        0 & e^{i(\omega+\lambda_B) t}}\right]\ ,
\eqno(A.5)
$$
where $\lambda_B={\lambda\over2}\big(1-{\mu\over\omega}\big)$ is precisely
the Berry phase contribution.

\vfill\eject

\centerline{\bf REFERENCES}
\bigskip\medskip

\item{1.} R. Alicki and K. Lendi, {\it Quantum Dynamical Semigroups and
Applications}, Lect. Notes Phys. {\bf 286}, (Springer-Verlag, Berlin, 1987)
\smallskip
\item{2.} V. Gorini, A. Frigerio, M. Verri, A. Kossakowski and
E.C.G. Surdarshan, Rep. Math. Phys. {\bf 13} (1978) 149
\smallskip
\item{3.} H. Spohn, Rev. Mod. Phys. {\bf 53} (1980) 569
\smallskip
\item{4.} A. Royer, Phys. Rev. Lett. {\bf 77} (1996) 3272
\smallskip
\item{5.} W.H. Louisell, {\it Quantum Statistical Properties of Radiation},
(Wiley, New York, 1973)
\smallskip
\item{6.} M.O. Scully and M.S. Zubairy,
{\it Quantum Optics} (Cambridge University Press, Cambridge, 1997)
\smallskip
\item{7.} C.W. Gardiner and P. Zoller,
{\it Quantum Noise}, 2nd. ed. (Springer, Berlin, 2000)
\smallskip
\item{8.} J. Ellis, N.E. Mavromatos and D.V. Nanopoulos, Phys. Lett.
{\bf B293} (1992) 37; Int. J. Mod. Phys. {\bf A11} (1996) 1489
\smallskip
\item{9.} F. Benatti and R. Floreanini, Ann. of Phys. {\bf 273} (1999) 58
\smallskip
\item{10.} S. Hawking, Comm. Math. Phys. {\bf 87} (1983) 395; Phys. Rev. D
{\bf 37} (1988) 904; Phys. Rev. D {\bf 53} (1996) 3099;
S. Hawking and C. Hunter, Phys. Rev. D {\bf 59} (1999) 044025
\smallskip
\item{11.} J. Ellis, J.S. Hagelin, D.V. Nanopoulos and M. Srednicki,
Nucl. Phys. {\bf B241} (1984) 381;
\smallskip
\item{12.} S. Coleman, Nucl. Phys. {\bf B307} (1988) 867
\smallskip
\item{13.} S.B. Giddings and A. Strominger, Nucl. Phys. {\bf B307} (1988) 854
\smallskip
\item{14.} M. Srednicki, Nucl. Phys. {\bf B410} (1993) 143
\smallskip
\item{15.} W.G. Unruh and R.M. Wald, Phys. Rev. D {\bf 52} (1995) 2176
\smallskip
\item{16.} L.J. Garay, Phys. Rev. Lett. {\bf 80} (1998) 2508;
Phys. Rev. D {\bf 58} (1998) 124015
\smallskip
\item{17.} For a presentation of these models, see: V.A. Rubakov,
{\it Phys. Usp.} {\bf 44} (2001) 871
\smallskip
\item{18.} F. Benatti and R. Floreanini, Nucl. Phys. {\bf B488} (1997) 335
\smallskip
\item{19.} F. Benatti and R. Floreanini, Phys. Lett. {\bf B401} (1997) 337
\smallskip
\item{20.} F. Benatti and R. Floreanini, Nucl. Phys. {\bf B511} (1998) 550
\smallskip
\item{21.} F. Benatti and R. Floreanini, Phys. Lett. {\bf B465} (1999) 260
\smallskip
\item{22.} F. Benatti, R. Floreanini and R. Romano,
Nucl. Phys. {\bf B602} (2001) 541
\smallskip
\item{23.} F. Benatti and R. Floreanini, Phys. Lett. {\bf B451} (1999) 422
\smallskip
\item{24.} F. Benatti and R. Floreanini, JHEP {\bf 02} (2000) 032
\smallskip
\item{25.} F. Benatti and R. Floreanini, Phys. Rev. D {\bf 64} (2001) 085015
\smallskip
\item{26.} F. Benatti and R. Floreanini, Phys. Rev. D {\bf 62} (2000) 125009
\smallskip
\item{27.} S. Carroll, G. Field and R. Jackiw, Phys. Rev. D {\bf 41}
(1990) 1231
\smallskip
\item{28.} M. Born and E. Wolf, {\it Principles of Optics},
(Pergamon Press, Oxford, 1980)
\smallskip
\item{29.} L.D. Landau and E.M. Lifshitz, {\it The Classical Theory of Fields},
(Pergamon Press, New York, 1975); {\it Quantum Mechanics},
(Pergamon Press, New York, 1975)
\smallskip
\item{30.} E. Collett, {\it Polarized Light}, (Marcel Dekker, New York, 1993)
\smallskip
\item{31.} C. Brosseau, {\it Fundamentals of Polarized Light},
(Wiley, New York, 1998)
\smallskip
\item{32.} E.B. Davies and H. Spohn, J. Stat. Phys. {\bf 19} (1978) 511
\smallskip
\item{33.} R. Alicki, J. Phys. A {\bf 12} (1979) L103
\smallskip
\item{34.} K. Lendi, Phys. Rev. A {\bf 33} (1986) 3358
\smallskip
\item{35.} F. Benatti and R. Floreanini,
Mod. Phys. Lett. {\bf A12} (1997) 1465;
Banach Center Publications, {\bf 43} (1998) 71;
Phys. Lett. {\bf B468} (1999) 287; Chaos, Solitons and Fractals
{\bf 12} (2001) 2631
\smallskip
\item{36.} A. Messiah, {\it Quantum Mechanics}, vol. II, (North-Holland,
Amsterdam, 1961)
\smallskip
\item{37.} M.V. Berry, Proc. R. Soc. Lond. A {\bf 392} (1984) 45
\smallskip
\item{38.} A. Shapere and F. Wilczek, {\it Geometric Phases in Physics},
(World Scientific, Singapore, 1989)
\smallskip
\item{39.} K. Lendi, J. Phys. {\bf A 20} (1987) 13
\smallskip
\item{40.} F. Benatti and R. Floreanini, Dissipative effects in the propagation
of polarized photons, in {\it Quantum Electrodynamics and Physics of 
the Vacuum},
G. Cantatore ed.,\break AIP Conference Proceedings, Vol. 564, 2001, p. 37
\smallskip
\item{41.} F. Benatti, R. Floreanini and A. Lapel, Cybernetics and Systems,
{\bf 32} (2001) 343

\bye